\documentclass[
superscriptaddress,
showpacs,
nofootinbib,
amsmath,amssymb,
pra,
twocolumn
]{revtex4-1}

\usepackage{graphicx}

\usepackage{braket}
\usepackage{natbib}
\usepackage{hyperref}
\usepackage{CJK}

\newcommand{\abs}[1]{\lvert#1\rvert}
\newcommand{\norm}[1]{\lVert#1\rVert}

\begin{document}

\begin{CJK*}{UTF8}{gbsn}
\title{On the relation between quantum Darwinism and approximate  quantum Markovianity}

\author{Xiao-Kan Guo~(郭肖侃)}
 \email{kankuohsiao@whu.edu.cn}
\affiliation{Department of Applied Mathematics, Yancheng Institute of Technology, Jiangsu 224051, China}
\author{Zhiqiang Huang (黄志强)}
\email{hzq@wipm.ac.cn}
\affiliation{State Key Laboratory of Magnetic Resonance and Atomic and Molecular Physics, Innovation Academy for Precision Measurement Science and Technology, Chinese Academy of Sciences, Wuhan 430071, China}
\date{\today}

\begin{abstract} 
There are strong evidences in the literature that quantum non-Markovianity would hinder the presence of Quantum Darwinism. 
 In this Letter, we study  the relation between quantum Darwinism and approximate quantum Markovianity for open quantum systems by exploiting the properties of quantum conditional mutual information. We show that for approximately Markovian quantum processes the conditional mutual information still has the scaling property for Quantum Darwinism. Then two general bounds on the backflow of information  are obtained, with which we can show that the presence of Quantum Darwinism restricts the information backflow and the quantum non-Markovianity must be small.
\end{abstract}


\maketitle

\section{Introduction}
Quantum decoherence gains importance  both in the foundational studies such as quantum measurements and in the practical studies such as the open quantum systems that are applicable to modern quantum technologies. An important recent development in quantum decoherence, known as Quantum Darwinism \cite{Z03,OPZ1,OPZ2,BZ05,BZ06,BZ08,QD,ZQZ09,ZQZ10,RZ10,RZ11,RZZ12,ZRZ14,ZRZ16,ZZ17,ULZZJ19}, explains how  classical objectivity emerges from the quantum formalism. Basically, in Quantum Darwinism, one treats the information transmission from the system $S$ to a fragment $F$ of the environment $E$ as a quantum channel, so that the accessible classical information as bounded by the Holevo information for different fragments $F_i$ of the environment $E$ can be compared. If these fragments agree upon the accessible information  to some 
reasonable extent, then we say the classical information thus obtained is objective. In this sense, Quantum Darwinism tells us how the environment witnesses the system in the process of quantum decoherence as an effective measurement channel.

As the Holevo bound involves an optimization procedure, it is difficult to calculate it. To circumvent this computational difficulty,  an alternative formulation of Quantum Darwinism is proposed in the recent works \cite{22a,22b,22c}, where one makes measurements on the fragments $F_i$ of $E$ and asks for the maximal information about $S$  that one can extract from these measurements. In contrast, in the original framework of Quantum Darwinism one makes measurements on the systems $S$ and considers the Holevo information defined by the conditional entropy conditioned on $S$. Such a change of measurements is reasonable in the sense that the two settings lead to the same scaling behavior of the Holevo information.
 It turns out that such a change of measurements not only makes the computations easier, but also reveals deep relations between the correlations in the environment $E$ and the information extracted from measuring $F_i$. For example, the Koashi-Winter monogamy relation can be invoked in this new framework to show the trade-off relation for the quantum correlations between $S$ and $F_i$ and the classical information of the $F_i$'s.

Interestingly, in this new framework of Quantum Darwinism, one can consider also the scaling of the conditional mutual information $I(S:F_l|F_k)$ \cite{22b}. This form of conditional mutual information also makes appearance in our recent work on quantum non-Markovianity \cite{HG21,HG22}. 
We therefore expect to study the connections between Quantum Darwinism and quantum (non-)Markovianity from a new perspective, in light of these new results. Previous studies on the connection between Quantum Darwinism and quantum (non-)Markovianity relies on specific models. In these model studies \cite{DM1,DM2,DM3,DM4,DM5,MS21}, there are strong evidences showing that  the presence of non-Markovianity would suppress Quantum Darwinism. However, there are  some models  \cite{DM7, OdD19,DM9} in which the  non-Markovian nature of  environment does not  prevent  the appearance of Quantum Darwinism. That the presence of non-Markovianity would hinder Quantum Darwinism is expected from the general understanding that quantum non-Markovianity is related to the information backflow from environment to system, but the absence of relations in several models calls upon further study in this direction.

In this Letter, we consider the relations between Quantum Darwinism and approximate   quantum Markovianity in a general setting. This is  operationally meaningful since the Holevo information decreases under  quantum operations and hence the quantum operations can only be operated approximately \cite{Cai03}.

We first use straightforward arguments from the approximately complete positive  maps \cite{BDW16} to explain in \S\ref{S3} that for  approximately Markovian processes the scaling property of Quantum Darwinism  still holds, meaning that for the quantum non-Markovianity to hinder Quantum Darwinism a large non-Markovianity quantifier should be required. In \S\ref{S4}, we study the backflow of correlation as quantified by the conditional mutual information and find such  backflow of correlation is upper bounded by both the classical information and the quantum discord. If the suppression as in \cite{22b} hold, then the backflow of information, which is one way of characterizing quantum non-Markovianity, is also greatly suppressed, thereby corroborating the approximately Markovian results discussed in \S\ref{S3}.

\section{Quantum Darwinism: Decohering the system or eavesdropping the environment}\label{S2}
We start   by reviewing briefly both the standard formulation of Quantum Darwinism \cite{Z03,OPZ1,OPZ2,BZ05,BZ06,BZ08,QD,ZQZ09,ZQZ10,RZ10,RZ11,RZZ12,ZRZ14,ZRZ16,ZZ17} and the recent formulation given in \cite{22a,22b,22c}. 

In Quantum Darwinism one considers the setting  of an open quantum system  $S$ interacting with an environment $E$ which consists of $N$ fragments (or subenvironments) $F_i$. The interaction between $S$ and $E$, or the quantum decoherence of $S$ by $E$, is considered as a quantum communication channel $\Lambda_{S\rightarrow F_i}$ such that information is transmitted from $S$ to $F_i$. Namely, correlation is created between $S$ and $F_i$ as the quantum  mutual information
\begin{equation}
I(S:F_i)=H_S+H_{F_i}-H_{SF_i}\neq0
\end{equation}
where $H_S=-\text{tr}\rho_S\log_2\rho_S$ is the von Neumann entropy.

To extract classical information about $S$ from $F_i$, one considers the quantum decoherence of   $S$ by projecting it to the pointer basis, $\{\ket{\hat{s}}\}$ \cite{Z03,PBasis}, and hence one obtains the Holevo information
\begin{equation}\label{2}
\chi(\hat{\Pi}_S:F_i)=H\Bigl(\sum_{\hat{s}}p_{\hat{s}}\rho_{F_i|\hat{s}}\Bigr)-\sum_{\hat{s}}p_{\hat{s}}H(\rho_{F_i|\hat{s}}),
\end{equation}
where $\hat{\Pi}_S=\sum_{\hat{s}}\pi_{\hat{s}}\ket{\hat{s}}\bra{\hat{s}}$ and $\rho_{F_i|\hat{s}}=\braket{\hat{s}|\rho_{SF_i}|\hat{s}}/p_{\hat{s}}$ is the fragment state conditioned on system's pointer state with the $p_{\hat{s}}$ being the probabilities for projecting to the pointer basis. The Holevo information is an upper bound on the classical  information transmittable in the quantum channel $\Lambda_{S\rightarrow F_i}$. For the successful extraction of information about $S$, one sets the condition
\begin{equation}\label{3}
\braket{\chi(\hat{\Pi}_S:F_i)}_{\# F_\delta}\approx(1-\delta) H_S
\end{equation}
where $\delta$ is a small number called the information deficit, and $\braket{}_{\# F_\delta}$  denotes the average over the fragments of size ${\# F_\delta}$.
Here, $H_S=H(\hat{\Pi}_S)$ quantifies the missing information about $S$ computed in the pointer basis, i.e. the classical information content of the pointer state when the quantum state is decohered. 
The condition \eqref{3} says that an observer only needs a number ${\# F_\delta}$ of fragments $F_i$ to retrieve $(1-\delta) H_S$ classical bits of information about $S$, and therefore the rest of the environment is redundant. Let $\# E=N$ be the size of the environment $E$, then we define the redundancy as
\begin{equation}
R_\delta=\frac{\# E}{\# F_\delta}.
\end{equation}
When the fragments $F_i$ are of the same size, we have $R_\delta=1/f_\delta$ with $f_\delta$ the fraction of the relevant fragments.

By using the complementary relation for the Holevo information and the quantum discord \cite{ZZ13}, we can rewrite the condition \eqref{3} as
\begin{equation}\label{5}
I(S:F_i)\approx(1-\delta) H_S.
\end{equation}
This is because the Holevo information is the upper bound on the transmittable classical information in the pointer state,
\begin{equation}\label{6}
\chi(\hat{\Pi}_S:F_i)=\max_{{s}}I({\Pi}_S:F_i)=J(\hat{\Pi}_S:F_i)
\end{equation}
where the maximum is taken over all possible eigenbasis, and
$J(\hat{\Pi}_S:F_i)$ is the classical correlation used in the definition of quantum discord. When the quantum correlation is small, we obtain \eqref{5}, with the understanding that it has been averaged, i.e. $I(S:F_i)=\braket{I(S:F_i)}_{\# F_\delta}$.

The most important property of $I(S:F_i)$ for Quantum Darwinism is  its {\it scaling property} under the increase of $\# F_\delta$. The study of various models confirms the general behavior of $I(S:F_i)$: At the initial stage, $I(S:F_i)$ follows a steep rise  with increasing $\# F_\delta$ as the classical correlation increases; then it saturates a long and flat plateau meaning that at this stage the added fragments are redundant, i.e. when 
\begin{equation}\label{7}
I(S:F_i)\geqslant(1-\delta) H_S;\end{equation}
finally, $I(S:F_i)$ follows again a steep rise to reach its maximum as the quantum correlation increases.

Now from \eqref{6} we see that it involves an optimization, which is generically hard. Also, since the Holevo information is only an upper bound, the above conditions for information retrieval can in fact be relaxed. To improve upon these limitations, the recent work \cite{22a} considers instead the measurements on $F_i$. Let us denote the part being measured by a check, e.g. the Holevo information in \eqref{2} becomes the asymmetric mutual information $J(\check{S}:F_i)$. When  $F_i$ is measured,  one has similarly 
\begin{equation}
J(S:\check{F}_i)=\chi(S:\check{F}_i),
\end{equation}
which is the maximal information about $S$ one can extract by measuring $F_i$. In fact, this $J(S:\check{F}_i)$ also upper bounds the classical accessible information \cite{22c}.
Numerical results  show that the scaling property for Quantum Darwinism still holds for $J(S:\check{F}_i)$. In particular, the plateau condition can be written as
\begin{equation}\label{9}
J(S:\check{F}_i)\geqslant(1-\delta) H_S.
\end{equation}

Using the symmetric mutual information $I(S:F_i)$ and the asymmetric mutual information $J(S:\check{F}_i)$, one can define the fragmentary discord
\begin{equation}\label{FDIS}
    D(S:\check{F}_i)=I(S:F_i)-J(S:\check{F}_i).
\end{equation}
 If we consider each fragment $F_i$ has its own information deficit $\delta_i$, and consider the averaged discord $\bar{D}(S:\check{F}_i)=\frac{1}{N}\sum_{i=1}^ND(S:\check{F}_i)$, then the following bound holds \cite{22b},
\begin{equation}\label{10}
\bar{D}(S:\check{F}_i)\leqslant\delta H_S
\end{equation}
where $\delta=\sum_{i=1}^N\delta_i/N$. Similarly, for the $\# F_\delta\leqslant\frac{N}{2}$ fragments, the bound can be rewritten as
\begin{equation}\label{11}
\bar{D}(S:\check{F}_i)\leqslant\bigl[1-(1-\delta)\frac{R_\delta}{N}\bigr] H_S.
\end{equation}
Clearly, since the information deficit is chosen to be small, the quantum correlation as quantified by the quantum discord is greatly suppressed. Importantly, the plateau condition \eqref{9} implies 
\begin{equation}\label{12}
I(S:F_l|F_k)\leqslant2\delta H_S,
\end{equation}
for $k\geqslant\#F_\delta,~k+l\leqslant N-\# F_\delta$. With \eqref{12}, one can directly test Quantum Darwinism by inspecting the scaling of the conditional mutual information $I(S:F_l|F_k)$, without facing the problem of optimization.
\section{Relation to approximately quantum Markovian processes}\label{S3}
We consider in this section the case in which the quantum channels involved deviate a little  from complete positivity. The complete positivity of the open quantum system dynamics is related to the quantum data processing inequality  \cite{Bus14}, and the deviation from the   quantum data processing inequality  indicates quantum non-Markovianity. A small deviation from the   quantum data processing inequality can give rise to an approximately complete positive map, and the converse is also true \cite{BDW16}.

To apply these deviation results to Quantum Darwinism, we consider the following novel scenario: Instead of focusing on the system $S$, we take the measurement made on the fragment $F_k$ as a quantum channel $\Lambda_{F_k\rightarrow F_k'}$. After the system $S$ has decohered and the information about $S$ has been transmitted to $F_k$, we assume that the fragment $F_k$ no longer interacts with $S$ but the interactions between different  $F_i$'s are still possible. With this assumption,
the quantum channel $\Lambda_{F_k\rightarrow F_k'}$ can be considered as an $F_k$-$F_l$ open system\footnote{In what follows we will often change the choices of system and environments. The meanings of the quantum channels should be read from the  contexts.}
where $F_k$ is the new open ``system'' of an observer's concern and another fragment $F_l$ with $l\neq k$ is the new environment for $F_k$. On the other hand,  the system $S$ becomes now a {\it steerable } extension of the $F_k$-$F_l$ open system, because the states $\rho_{F_kF_l}$ can be engineered according to the pointer basis of $S$. This way, we have adapted the new setting of Quantum Darwinism to that of \cite{Bus14}.

Now suppose that the   quantum channel $\Lambda_{F_k\rightarrow F_k'}$  is approximately CPTP, in the sense that 
\begin{equation}
\frac{1}{2}\norm{\sigma_{SF'_k}-\Lambda_{F_k\rightarrow F_k'}(\rho_{SF_k})}_1\leqslant\epsilon
\end{equation}
where $\epsilon\in[0,1]$ and $\sigma_{SF'_k}=\text{Tr}_{F_1}\sigma_{SF_kF_l}$ with 
$\sigma_{SF_kF_l}=U_{F_kF_l}(\rho_{SF_kF_l})U^\dag_{F_kF_l}$ being the partially evolved state. 
In this context, we can consider the mutual information calculated with respect to the state $\sigma_{SF'_k}$,
\begin{equation}
I(S:F'_k)_\sigma=H(S)_\sigma-H(S|F'_k)_\sigma
\end{equation}
between the original system $S$ and the fragment $F'_k$ resulted from the quantum channel $\Lambda_{F_k\rightarrow F_k'}$. Since the original system $S$ does not evolve under $\Lambda_{F_k\rightarrow F_k'}$, we have $\sigma_{SF'_k}=\Lambda_{F_k\rightarrow F_k'}(\rho_{SF_k})$.
Then by the Alicki-Fannes-Winter inequality for conditional entropies \cite{AF04,W16}, 
\begin{equation}\label{AFW}
\abs{H(S|F'_k)_\rho-H(S|F'_k)_\sigma}\leqslant2\epsilon\log\abs{S}+(1+\epsilon)h[\frac{\epsilon}{1+\epsilon}],
\end{equation}
where $\abs{S}$ is the dimension of the  Hilbert space for the original system $S$ and
 $h[x]=-x\log x-(1-x)\log(1-x)$ is the binary entropy, we have 
 \begin{align}
& I(S:F'_k)_\sigma=H(S)_{\Lambda(\rho)}-H(S|F'_k)_\sigma\nonumber\\
 \leqslant&H(S)_{\Lambda(\rho)}-H(S|F'_k)_{\Lambda(\rho)}+2\epsilon\log\abs{S}+(1+\epsilon)h[\frac{\epsilon}{1+\epsilon}]\nonumber\\
 =&I(S:F'_k)_{\Lambda(\rho)}+2\epsilon\log\abs{S}+(1+\epsilon)h[\frac{\epsilon}{1+\epsilon}]\nonumber\\
 \leqslant&I(S:F_k)_{\rho}+2\epsilon\log\abs{S}+(1+\epsilon)h[\frac{\epsilon}{1+\epsilon}]\label{177}
 \end{align}
 where the first inequality is due to  the Alicki-Fannes-Winter inequality \eqref{AFW} and the second inequality comes from the data processing inequality for the  $F_k$-$F_l$ open system. 
 Furthermore,  by moving $I(S:F'_k)_{\rho}$ in \eqref{177} to the left-hand side of the inequality, we see that two von Neumann entropy terms cancel and a difference of two relative entropies remains. Since any other fragment $F_l$ with $l\neq k$ also does not evolve under $\Lambda_{F_k\rightarrow F_k'}$, we can write $F'_k=F_kF_l$, 
 so that the difference becomes
 \begin{equation}
 H(S:F_k)_\rho-H(S:F_kF_l)_\rho=I(S:F_l|F_k)_\rho
 \end{equation}
which  is exact a conditional mutual information. Thus, 
 we  have the  bound on the  conditional mutual information
\begin{equation}\label{15}
I(S:F_l|F_k)_\rho\leqslant2\epsilon\log\abs{S}+(1+\epsilon)h[\frac{\epsilon}{1+\epsilon}].
\end{equation}
The above inequalities  first appeared in Theorem 8 of \cite{BDW16}.

The bound in \eqref{15} has the same qualitative structure as \eqref{12}: two bounds both depend on a small parameter and  the ``entropy'' of $S$.\footnote{In \eqref{15}, the term $\log\abs{S}$ can be roughly understood as the statistical-mechanical entropy in the Boltzmann sense.} 
We therefore see that, with a small deviation from complete positivity, the scaling behavior \eqref{12}  of conditional mutual information for Quantum Darwinism {\it  can still hold}, if we choose proper parameters $\epsilon,\delta$, and
 assume that the fragment $F_k$ no longer interacts with the system $S$. In other words, the  approximately Markovian quantum processes will not hinder the presence of Quantum Darwinism. This understanding is consistent with the strict  bounds on the deviation from the quantum Markovianity  \cite{FR15,FMP19}.
 
Next, let us consider the Holevo information of the ensemble of states $\mathcal{S}=\{p_{\hat{s}},\rho_{\hat{s}}\}$,
\begin{equation}
\chi(\mathcal{S})=H\Bigl(\sum_{\hat{s}}p_{\hat{s}}\rho_{\hat{s}}\Bigr)-\sum_{\hat{s}}p_{\hat{s}}H(\rho_{\hat{s}}).
\end{equation}
Then the Holevo information \eqref{2} of the quantum channel $\Lambda_{S\rightarrow F_i}$ is the Holevo information $\chi(\mathcal{F})$ of the output ensemble $\mathcal{F}=\{p_{\hat{s}},\rho_{F_i|\hat{s}}\}$. The monotonicity of the Holevo information under quantum channels gives $\chi(\mathcal{S})\geqslant\chi(\mathcal{F})$, and when the equality holds there exists a (Petz) recovery channel $\Phi_{F_i\rightarrow S}$ such that $\Phi_{F_i\rightarrow S}\circ\Lambda_{S\rightarrow F_i}(\rho_{\hat{s}})=\rho_{\hat{s}}$ \cite{HJPW04}. The existence of recovery channel is equivalent to the  backflow of {\it all} information from $F_i$ to $S$, and we see that this requires the Holevo information of the ensemble of states is intact under the quantum channel $\Lambda_{S\rightarrow F_i}$, i.e. 
\begin{equation}
\chi(\mathcal{S})=\chi(\mathcal{F})=\chi(\hat{\Pi}_S:F_i).\end{equation}
Small deviations from this equality can be bounded in the following way (cf. Theorem 6 of \cite{BDW16}). 

We consider an auxiliary classical system $X$ denoting the measurement outcomes $x$,
 such that the ensembles of states $\mathcal{S}$ and $\mathcal{F}$ can be represented by the following classical-quantum states,
\begin{align}
\rho_{x\hat{s}}=&\sum_xp_x\ket{x}\bra{x}\otimes \rho_{\hat{s}},\nonumber\\
\sigma_{xF_i|\hat{s}}=&\sum_xp_x\ket{x}\bra{x}\otimes \Lambda_{S\rightarrow F_i}(\rho_{\hat{s}}).
\end{align}
The quantum discord for a classical-quantum state vanishes, so the Hovelo information is just the mutual information. Hence we have
\begin{align}
&\chi(\mathcal{S})-\chi(\mathcal{F})=I(X:S)_\rho-I(X:F)_\sigma\nonumber\\
=&D(\rho_{x\hat{s}}||I_X\otimes\rho_A)-D({\bf1}\otimes\Lambda(\rho_{x\hat{s}})||{\bf1}\otimes\Lambda(I_X\otimes\rho_A))
\end{align}
where $D(\rho||\sigma)$ is the quantum relative entropy.
 By the improved monotonicity of quantum relative entropies in the presence of recovery channel $\Phi$ \cite{JRSWW18},
\begin{equation}
D(\rho||\sigma)\geqslant D(\Lambda(\rho)||\Lambda(\sigma))-\log F(\rho, \Phi\circ\Lambda(\rho)),
\end{equation}
where $F(\rho,\sigma)=\norm{\sqrt{\rho}\sqrt{\sigma}}_1^2$ is the fidelity,
we get
\begin{align}
\chi(\mathcal{S})-\chi(\mathcal{F})\geqslant&-\log F(\rho_{x\hat{s}}, ({\bf1}\otimes\Phi)\circ({\bf1}\otimes\Lambda)(\rho))\nonumber\\
=&-2\log\sum_{\hat{s}}p_{\hat{s}}\sqrt{F(\rho_{\hat{s}},\Phi\circ\Lambda(\rho_{\hat{s}}))}.\label{18}
\end{align}
The final equality in \eqref{18} is due to the particular structure of classical-quantum states.

 The inequality \eqref{18} is the constraint for the existence of the approximate recovery channel or approximately full information backflow. 
As far as the quantum decoherence is concerned, we expect a large fidelity $F(\rho_{\hat{s}},\Phi\circ\Lambda(\rho_{\hat{s}}))$, 
since the decoherence features an irreversible information transmission. 
However, when strong non-Markovianity is present, it is still  possible for such an approximate recovery channel to exist, in which case there is no emergence of classicality as the distinguishability of quantum states quantified by the Hovelo information is approximately preserved. Therefore, to hinder Quantum Darwinism requires strong non-Markovianity.

\section{Relation to backflow of correlation}\label{S4}
It is well-known by now that  the quantum non-Markovianity  of open quantum dynamics can be quantified by the information backflow  from environment to system, or the increase in distinguishability of system's states \cite{BLP09}. Recently, we have shown in \cite{HG21} that the backflow of quantum mutual information can be equivalently reformulated as the backflow of conditional mutual information in a system-environment-ancilla setting. Here, we shall consider the backflow of conditional mutual information in the new setting of Quantum Darwinism.

We consider first the pure state $\rho_{SE}$
for the system-environment total system. Since for pure states both the quantum discord and the classical correlation takes the maximal value of $H_S$, we have 
\begin{equation}
I(S:E)=D(S:\check{E})+J(S:\check{E})=2H_S.
\end{equation}
In the system-environment-ancilla setting of \cite{HG21}, the ancilla-system initial state is maximally entangled, and the joint state is
\begin{equation}
   \Bigl(\sum_{i}\frac{1}{\sqrt{N_{S}}} \ket{i}_A\otimes\ket{i}_S\Bigr)\otimes \ket{\phi}_E.
\end{equation}
In this case, $I(A:S)_{\rho_i}=2H_A$. In other words, initially the ancillary $A$ possesses all the information about $S$ (and vice versa). Notice that the ancillary $A$ does not evolve, so in this setting, the ancillary system A is used only to keep track of the information flow. Since the initial environment contains no information about the initial system, we have $I(A:E|S)=0$. After the information of $S$ is spread into $E$ with unitary evolution $U^{SE}$, we have in general
$I(A:S)\neq2H_A$ because of the data processing inequality, $I(A:SE)=2H_A$ as $A$ does not evolve under $U_{SE}$, and hence $I(A:E|S)\neq0$. Therefore, the memory effect can be reflected by the non-monotonicity of the  conditional mutual information $I(A:E|S)=I(A:SE)-I(A:S)$, and the upper bound of the backflow is $I(A:E|S)$. 

Since the speed of information backflow is expected to be finite, we usually do not need to consider the information backflow from the whole environment $E$, but we can consider the information backflow from the subenvironment $E_\text{sub}$ near the system $S$. So we can study $I(A:E_\text{sub}|S)$ to bound information backflow.

Notice that  the ancillary $A$ can be considered either as a fragment $F_i$ of the environment or  a special observer who makes measurements on $S$. 
We first consider the former interpretation of $A$ as fragment of the joint environment $EA$.
To avoid confusions, let us label the   $(S^*,E^*,A^*)$ in the system-environment-ancilla setting with a $*$.
We can take the $A^*$ as the system $S$ in Quantum Darwinism setting and take $S^*$ as a marked fragment $F_1$ in Quantum Darwinism setting. Let us consider the following
\begin{equation}
I(A^*:E^*_\text{sub}|S^*)\equiv I(S:{F}_l|{F}_1),
\end{equation}
where the RHS is understood as the information transmission from $S$ to $F_1$ in the setting of Quantum Darwinism, which can be alternatively understood as in the LHS as information flow from $A^*$ to $S^*$. 
If we adopt the interpretation of $A^*$ as a fragment of the joint environment $E^*A^*$, this information transmission is the backflow of information. 
So we have\footnote{Here the notation $F_{l+1}$ means there are $l+1$ fragments.}
    \begin{align}
        &I(A^*:E^*_{sub}|S^*)= I(S:{F}_l|{F}_1)\notag \\
        =&I(S:F_{l+1})-I(S:F_1)\notag \\
        =&I(S:F_{\#F_\delta})-I(S:F_1)+I(S:F_{l+1})-I(S:F_{\#F_\delta})\notag \\
        =&I(S:F_{\#F_\delta})-I(S:F_1)+I(S:{F}_{l-\#F_\delta+1}|{F}_{\#F_\delta})\notag \\
        \leqslant&I(S:F_{\#F_\delta})-I(S:F_1)+2\delta H_S \label{20}\\
        \approx&J(S;F_{\#F_\delta})+D(S;F_{\#F_\delta})-(1-2\delta-\delta')H(S)\label{21}
    \end{align}
where in \eqref{20} we have used the scaling \eqref{12}, and in \eqref{21} we have used \eqref{5}, \eqref{FDIS}  and $\delta'$ is the information deficit for this particular case, defined by $I(S:F_1)\approx(1-\delta')H_S$. That is, when $A^*$ keeps almost all the classical information about $S^*$, $\delta'$ will be small.  
 By expressing \eqref{5} in terms of $J$,  we obtain
\begin{equation}
I(A^*:E^*_\text{sub}|S^*)\lessapprox(\delta'+\delta)H_S +D(S;F_{\#F_\delta})\label{22}.
\end{equation}
This result \eqref{22} shows that the conditional mutual information $I(A^*:E^*_\text{sub}|S^*)$ is  bounded both by the classical information  and the quantum correlation. This clearly shows that  the backflow of information contains both classical and quantum correlations. Now, in the  parameter range $ l\leqslant N-2\#F_\delta$, by further using the bound \eqref{11}, we see that the conditional mutual information, which is to be backflowed from $A^*$ to $S^*$, is greatly suppressed. 

Notice that the backflow of information from $A^*$ to $S^*$ is understood by the standard setting of Quantum Darwinism as the channel from $S$ to $F_i$. Although a little strange, this is allowable by the general results on the emergence of Quantum Darwinism in a general many-body quantum system \cite{BPH15,KTPA18,CLAT21,QR21,RZZ16,Rie17,Oll22}.  In a sense, we have modeled the emergence of Quantum Darwinism from $A^*$ to $\{E_{sub}^*\}_{i}\cup\{S^*\}$.

Next, we consider an alternative setting by taking $A$ as an observer who measures $S$. Suppose the initial system-environment state  is
    \begin{equation}\label{SODAW0}
        \ket{\psi_{SE}(0)}=\Bigl(\sum_{{s}} \ket{{s}}_S\Bigr)\otimes \ket{\phi}_E
    \end{equation}
where the $\ket{s}$ is the eigenbasis of $S$ and the normalization is hidden. 
After $A$ measured $S$ by projections $\ket{\hat{s}}\bra{\hat{s}}$ to the pointer basis $\ket{\hat{s}}$, the  joint state is then
    \begin{equation}\label{SOH0}
        \ket{\psi'_{ASE}(0)}=\Bigl(\sum_{\hat{s}} \sqrt{P_{\hat{s}}}\ket{\hat{s}}_A\otimes\ket{\hat{s}}_S\Bigr)\otimes \ket{\phi}_E
    \end{equation}
 where $\sqrt{P_{\hat{s}}}=\braket{\hat{s}|s}$. In this case, the $AS$-state is  entangled, and the observer $A$ keeps all the classical information about $S$. We have in fact  returned to the system-environment-ancilla setting of \cite{HG21}.
After the time evolution of $SE$ by $U^{SE}$ (or under the quantum channel $\Lambda_{S\rightarrow E}$), we have 
 \begin{equation}\label{SOHP}
    \ket{\psi'_{ASE}}=U^{SE}  \ket{\psi'_{ASE}(0)}=\sum_{\hat{s}} \sqrt{P_{\hat{s}}}\ket{\hat{s}}_A\otimes\ket{\hat{s}}_S\otimes \ket{\phi_{\hat{s}}}_E.
\end{equation}
The $SE$-part is now a branching state
\begin{equation}\label{SOH}
    \ket{\psi_{SE}}=U^{SE}  \ket{\psi_{SE}(0)}=\sum_{\hat{s}} \sqrt{P_{\hat{s}}}\ket{\hat{s}}_S\otimes \ket{\phi_{\hat{s}}}_E.
\end{equation}

To proceed, we observe that joint state $\ket{\psi'_{ASE}}$ can be obtained from the state $\ket{0}_A\otimes\ket{\psi_{SE}}$, with $\ket{0}_A$ a pure state, by a unitary $U^{AS}: \ket{\hat{s}0}\to \ket{\hat{s}\hat{s}}$,
\begin{equation}
    \ket{\psi'_{ASE}}=U^{AS} \ket{0}_A\otimes \ket{\psi_{SE}}.
\end{equation}
Since the pure state has zero von Neumann entropy and the unitary does not increase von Neumann entropy, we have in this case,
\begin{equation}\label{DCMI1}
    I(AS:E_\text{sub})_{\ket{\psi'_{ASE}}}=  I(S:E_\text{sub})_{ \ket{\psi_{SE}}}.
\end{equation}
Let us take partial trace of \eqref{SOHP}
\begin{equation}\label{27}
    \text{Tr}_{A\overline{E}_\text{sub}} \Pi_{\psi'}^{ASE}=\sum_{\hat{s}} P_{\hat{s}} \Pi_{\hat{s}}^S \otimes  ( \text{Tr}_{\overline{E}_\text{sub}} \Pi_{\phi_{\hat{s}}}^{E}),
\end{equation}
where $\overline{E}_\text{sub}E_\text{sub}=E$.
Now we can assume the condition of  good decoherence \cite{ZQZ10}  that the
off-diagonal terms of $\rho_{S}$  are negligible, so that  \eqref{27} is a classical-quantum state with zero quantum discord, i.e.
    \begin{equation}\label{DCMI2}
        I(S:E_\text{sub})_{\ket{\psi'_{ASE}}}=  \chi(\check{S}:E_\text{sub})_{ \ket{\psi_{SE}}}.
    \end{equation}
Finally, combining \eqref{DCMI1} and \eqref{DCMI2}, we obtain
    \begin{align}
        I(A:E_\text{sub}|S)_{\ket{\psi'}}&=I(AS:E_\text{sub})_{\ket{\psi'}}-I(S:E_\text{sub})_{\ket{\psi'}}\notag \\
       & =I(S:\mathcal{F}_l)- \chi(\check{S}:\mathcal{F}_l)=D(\check{S}:\mathcal{F}_l).\label{29}
    \end{align}
Since we have assumed the good decoherence, we have indeed $D(\check{S}:\mathcal{F}_l)\approx0$ \cite{22a}.

In the present case, the $I(A:E_\text{sub}|S)_{\ket{\psi'}}$ contains the part that can be possibly  backflowed, and any backflow of conditional mutual information is also bounded as above. Note that, in comparison to \eqref{22}, the classical information is not present in the bound \eqref{29}; this is because the interpretations of $A$ differ in two cases. Here, $A$ contains all the classical information (i.e. the pointer observables) of $S$, while for \eqref{22} this classical information needs to be transmitted.

\section{Conclusion}\label{IV}
We have discussed the relation between Quantum Darwinism and approximate quantum non-Markovianity. In \S\ref{S3}, we have showed that given an approximately Markovian quantum process with an approximate CPTP map, the conditional mutual information still satisfies the scaling property for Quantum Darwinism as is recently given in \cite{22b}. While in \S\ref{S4}, we have showed the converse that the existence of Quantum Darwinism greatly suppresses the backflow of correlation. In this sense, we have showed the following {\it physical} statement:
\begin{quote}
{\it Quantum Darwinism is present in an open quantum system if and only if the open system is Markovian or approximately Markovian. }
\end{quote}
A caveat is that, these two-way restrictions are obtained by admitting several assumptions. For example,  in \S\ref{S3} we have assumed that the measured fragment no longer interacts with the original system, and in  \S\ref{S4} we also assume that the subenvironment/fragment near the system contains almost all the classical information, which is equivalent to say that the  plateau of mutual information needs to be very long and flat. In addition, we also implicitly work with a simple scenario where the fragments of environments are independent\footnote{Indeed, in the very recent work \cite{CIP23}, the nonlocal properties of the fragments of the environments are shown to be important to distinguish between Quantum Darwinism and the spectrum broadcasting structure.} and the information spreading is uniform. If these assumptions do not hold in some situations, the relation between Quantum Darwinism and approximate quantum non-Markovianity obtained in this Letter might fail. We look forward to further consider the physical necessity of these assumptions.

This result of this Letter is consistent with the numerical works \cite{DM1,DM2,DM3,DM4,DM5,MS21} where
the highest redundancy is reached when the dynamics is fully Markovian,  the redundancy tends to
zero when the dynamics is non-Markovian. (Particularly, \cite{MS21} also considers the information flow from the environment-point-of-view.)
With the explicit bounds obtained in this paper, we can further understand the boundary between these two regimes. 
But the numerical results supporting the irrelevance between Quantum Darwinism and non-Markovianity \cite{DM7, OdD19,DM9} cannot be understood easily from the point of view of present work. These  irrelevance results involve other aspects of classical objectivity such as the spectrum broadcast structures \cite{SBS}, which is worthy of further investigations.

\begin{acknowledgments}
We would like to thank Diogo O. Soares-Pinto for helpful discussions. 
ZH  is supported by the National Natural Science Foundation of China under Grant Nos. 12305035.
\end{acknowledgments}

\end{CJK*}


\begin{thebibliography}{99}
\bibitem{Z03} W. H. Zurek, Decoherence, einselection, and the quantum origins of the classical, Rev. Mod. Phys. {\bf75}, 715 (2003).
\bibitem{OPZ1} H. Ollivier, D. Poulin, and W. H. Zurek, Objective properties from subjective quantum states: Environment as a witness, Phys. Rev. Lett. {\bf93}, 220401 (2004).
\bibitem{OPZ2} H. Ollivier, D. Poulin, and W. H. Zurek, Environment as a witness: Selective proliferation of information and emergence of objectivity in a quantum universe, Phys. Rev. A {\bf72}, 042113 (2005).
\bibitem{BZ05} R. Blume-Kohout and W. H. Zurek, A simple example of ``Quantum Darwinism'': Redundant information storage in many-spin environments, Found. Phys. {\bf35}, 1857 (2005).
\bibitem{BZ06} R. Blume-Kohout and W. H. Zurek, Quantum Darwinism: Entanglement, branches, and the emergent classicality of redundantly stored quantum information, Phys. Rev. A {\bf73}, 062310 (2006).
\bibitem{BZ08} R. Blume-Kohout and W. H. Zurek, Quantum Darwinism in quantum Brownian motion, Phys. Rev. Lett. {\bf101}, 240405 (2008).
\bibitem{QD} W. H. Zurek, Quantum Darwinism, Nature Phys. {\bf5}, 181 (2009).
\bibitem{ZQZ09} M. Zwolak, H.-T. Quan, and W. H. Zurek, Quantum Darwinism in a mixed environment, Phys. Rev. Lett. {\bf103}, 110402 (2009).
\bibitem{ZQZ10} M. Zwolak, H.-T. Quan, and W. H. Zurek, Redundant imprinting of information in nonideal environments: Objective reality via a noisy channel, Phys. Rev. A {\bf81}, 062110 (2010).
\bibitem{RZ10}  C. J. Riedel and W. H. Zurek, Quantum Darwinism in an everyday environment: Huge redundancy in scattered photons, Phys. Rev. Lett. {\bf105}, 020404 (2010).
\bibitem{RZ11}  C. J. Riedel and W. H. Zurek, Redundant information from thermal illumination:
Quantum Darwinism in scattered photons, New J. Phys. {\bf13}, 073038 (2011).
\bibitem{RZZ12} C. J. Riedel, W. H. Zurek, and M. Zwolak, The rise and fall of redundancy in decoherence
and quantum Darwinism, New J. Phys. {\bf14}, 083010 (2012).
\bibitem{ZRZ14} M. Zwolak, C. J. Riedel, and W. H. Zurek, Amplification, redundancy, and quantum Chernoff information, Phys. Rev. Lett. {\bf112}, 140406 (2014).
\bibitem{ZRZ16} M. Zwolak, C. J. Riedel, and W. H. Zurek, Amplification, decoherence, and the acquisition of information by spin environments, Sci. Rep. {\bf6}, 25277 (2016).
\bibitem{ZZ17} M, Zwolak and W. H. Zurek, Redundancy of einselected information in quantum Darwinism:
The irrelevance of irrelevant environment bits, Phys. Rev. A {\bf95}, 030101(R) (2017).
\bibitem{ULZZJ19} T. K. Unden, D. Louzon, M. Zwolak, W. H. Zurek, and F. Jelezko, Revealing the emergence of classicality using nitrogen-vacancy centers, Phys. Rev. Lett. {\bf123}, 140402 (2019).

\bibitem{22a} A. Touil, B. Yan, D. Girolami, S. Deffner, and W. H. Zurek,  Eavesdropping on the decohering environment: Quantum Darwinism, amplification, and the origin of objective classical reality, Phys. Rev. Lett. {\bf128}, 010401 (2022).
\bibitem{22b} D. Girolami, A. Touil, B. Yan, S. Deffner, and W. H. Zurek, Redundantly amplified information suppresses quantum correlations in many-body systems, Phys. Rev. Lett. {\bf129}, 010401 (2022).
\bibitem{22c} M. Zwolak, Amplification, inference, and the manifestation of objective classical information, Entropy {\bf24}, 781 (2022).
\bibitem{HG21} Z.-Q. Huang and X.-K. Guo, Quantifying non-Markovianity via conditional mutual information, Phys. Rev. A {\bf104}, 032212 (2021).
\bibitem{HG22} Z.-Q. Huang and X.-K. Guo, Classical and quantum parts of conditional mutual information
for open quantum systems, Phys. Rev. A {\bf106}, 042412 (2022).
\bibitem{DM1} G. L. Giorgi, F. Galve, and R. Zambrini, Quantum Darwinism and non-Markovian dissipative dynamics from quantum phases of the spin-$1/2$ XX model, Phys. Rev. A {\bf92}, 022105 (2015).
\bibitem{DM2} F. Galve, R. Zambrini, and  S. Maniscalco, Non-Markovianity hinders Quantum Darwinism, Sci. Rep. {\bf6}, 19607 (2016).
\bibitem{DM3} G. Pleasance and B. M. Garraway, Application of quantum Darwinism to a structured environment, Phys. Rev. A {\bf96}, 062105 (2017).
\bibitem{DM4} N. Milazzo, S. Lorenzo, M. Paternostro, and G. M. Palma, Role of information backflow in the emergence of quantum Darwinism, Phys. Rev. A {\bf100}, 012101 (2019).
\bibitem{DM5} S. Lorenzo, M. Paternostro, and G. M. Palma, Reading a qubit quantum state with a quantum
meter: Time unfolding of quantum darwinism and quantum information flux, Open Syst. Inf. Dyn. {\bf26}, 1950023 (2019).
\bibitem{MS21} W. S. Martins and D. O. Soares-Pinto, Suppressing information storage in a structured thermal bath: Objectivity and non-Markovianity, arXiv:2110.03490.
\bibitem{DM7} A. Lampo, J. Tuziemski, M. Lewenstein, and J. K. Korbicz, Objectivity in the non-Markovian spin-boson model, Phys. Rev. A {\bf96},  012120 (2017).
\bibitem{OdD19} S. M. Oliveira, A. L. de Paula Jr., and R. C. Drumond, Quantum Darwinism and non-Markovianity in a model of quantum harmonic oscillators, Phys. Rev. A {\bf100}, 052110 (2019).
\bibitem{DM9} N. Megier, A. Smirne, S. Campbell, and B. Vacchini, Correlations, information backflow, and objectivity in a class of pure dephasing models, Entropy  {\bf24}, 304 (2022).
\bibitem{Cai03} Q.-y. Cai, Accessible information and quantum operations, arXiv:quant-ph/0303117.
\bibitem{BDW16} F. Buscemi, S. Das, and M. M. Wilde, Approximate reversibility in the context of entropy gain, information gain, and complete positivity, Phys. Rev. A {\bf93}, 062314 (2016).




\bibitem{PBasis} W. H. Zurek, Pointer basis of quantum apparatus: Into what mixture does the wave packet collapse? Phys. Rev. D {\bf24}, 1516 (1981).
\bibitem{ZZ13} M. Zwolak and W. H. Zurek, Complementarity of quantum discord
and classically accessible information, Sci. Rep. {\bf3}, 1729 (2013).

\bibitem{Bus14} F. Buscemi, Complete positivity, Markovianity, and the quantum data-processing inequality, in the presence of initial system-environment correlations, Phys. Rev. Lett. {\bf113}, 140502 (2014).

\bibitem{AF04} R. Alicki and M. Fannes, Continuity of quantum conditional information, J. Phys. A: Math. Gen. {\bf37}, L55 (2004).
\bibitem{W16} A. Winter, Tight uniform continuity bounds for quantum entropies: Conditional entropy, relative entropy distance and energy constraints, Commun. Math. Phys. {\bf347}, 291 (2016).
\bibitem{FR15} O. Fawzi and R. Renner, Quantum conditional mutual information and approximate Markov chains, Commun. Math. Phys. {\bf340}, 575 (2015).
\bibitem{FMP19} P. Figueroa-Romero, K. Modi, and F. A. Pollock, Almost Markovian processes from closed dynamics, Quantum {\bf3}, 136 (2019).
\bibitem{HJPW04} P. Hayden, R. Jozsa, D. Petz, and A. Winter, Structure of states which satisfy strong subadditivity of quantum entropy with equality, Commun. Math. Phys. {\bf246}, 359 (2004).
\bibitem{JRSWW18} M. Junge, R. Renner, D. Sutter, M. M. Wilde, and A. Winter, Universal recovery maps and approximate sufficiency of quantum relative entropy, Ann. Henri Poincar\'e {\bf19}, 2955 (2018).
\bibitem{BLP09} H.-P. Breuer, E.-M. Laine, and J. Piilo, Measure for the degree of non-Markovian behavior of quantum processes in open systems, Phys. Rev. Lett. {\bf103}, 210401 (2009).

\bibitem{BPH15} F. G. S. L. Brandao, M. Piani, and P. Horodecki, Generic emergence of classical features in quantum Darwinism, Nature Commun. {\bf6}, 7908 (2015).
\bibitem{KTPA18} P. A. Knott, T. Tufarelli, M. Piani, and G. Adesso, Generic emergence of objectivity of observables in infinite dimensions, Phys. Rev. Lett. {\bf121}, 160401 (2018).
\bibitem{CLAT21} E. Colafranceschi, L. Lami, G. Adesso, and T. Tufarelli, Refined diamond norm bounds on the emergence of objectivity of observables, J. Phys. A: Math. Theor. {\bf53}, 395305 (2020).
\bibitem{QR21} X.-L. Qi and D. Ranard, Emergent classicality in general multipartite states and channels, Quantum {\bf5}, 555 (2021).

\bibitem{RZZ16} C. J. Riedel, W. H. Zurek, and M. Zwolak, Objective past of a quantum universe: Redundant records of consistent histories,  Phys. Rev. A {\bf93}, 032126 (2016).
\bibitem{Rie17} C. J. Riedel, Classical branch structure from spatial redundancy in a many-body wave function, Phys. Rev. Lett. {\bf118}, 120402 (2017).
\bibitem{Oll22} H. Ollivier, Emergence of objectivity for quantum many-body systems, Entropy {\bf24}, 277 (2022).
\bibitem{CIP23} D. A. Chisholm, L. Innocenti, and G. M. Palma, The meaning of redundancy and consensus in quantum objectivity, Quantum {\bf7}, 1074 (2023).
\bibitem{SBS} J. K. Korbicz, Roads to objectivity: Quantum Darwinism, spectrum broadcast structures,
and strong quantum Darwinism-a review, Quantum {\bf5}, 571 (2021).


\end{thebibliography}
\end{document}